\documentclass[twocolumn,showpacs,preprintnumbers,amsmath,amssymb,superscriptaddress]{revtex4}
\usepackage{graphicx}
\usepackage{color}
\usepackage{subfigure}

\begin{document}

\title{Phase transitions of quasistationary states in the Hamiltonian Mean Field model}

\author{Pierre de Buyl}
\affiliation{Center for Nonlinear Phenomena and Complex Systems,
Universit{\'e} Libre de Bruxelles, Code Postal 231, Campus Plaine, B-1050 Brussels, Belgium}
\author{Duccio Fanelli}
\affiliation{Dipartimento di Energetica ``S. Stecco'' and CSDC, University of Florence, 
CNISM and INFN, Via S. Marta 3, 50139 Florence, Italy}
\author{Stefano Ruffo}
\affiliation{Dipartimento di Energetica ``S. Stecco'' and CSDC, University of Florence, 
CNISM and INFN, Via S. Marta 3, 50139 Florence, Italy}
\affiliation{Laboratoire de Physique de l'{\'E}cole Normale Sup{\'e}rieure de Lyon,
Universit{\'e} de Lyon, CNRS, 46 All{\'e}e d'Italie, 69364 Lyon c{\'e}dex 07, France}

\date{\today}

\begin{abstract}
The out-of equilibrium dynamics of the Hamiltonian Mean Field (HMF) model is studied in presence of an 
externally imposed magnetic field $h$. Lynden-Bell's theory of violent relaxation is revisited and shown to 
adequately capture the system dynamics, as revealed by direct Vlasov based numerical simulations in the limit of 
vanishing field. This includes the existence of an out-of-equilibrium phase transition separating 
magnetized and non magnetized phases. We also monitor the fluctuations in time of the magnetization,
which allows us to elaborate on the choice of the correct order parameter when challenging the
performance of Lynden-Bell's theory. The presence of the field $h$ removes the phase transition, as 
it happens at equilibrium. Moreover, regions with negative susceptibility are numerically 
found to occur, in agreement with the predictions of the theory.
\end{abstract}

\maketitle

Long-range interacting systems are characterized by a slowly decaying
interparticle potential, which in fact results in a substantial degree of
coupling among far away components. In these systems, energy is consequently
non-additive and this fact yields a large gallery of peculiar, apparently
unintuitive, phenomena: the specific heat can be negative in
the microcanonical ensemble, and temperature jumps may appear at microcanonical
first-order phase transitions~\cite{reviewRuffo,Leshouches}. Canonical and microcanonical statistical
ensembles can therefore be non-equivalent in presence of long-range
interactions, an intriguing possibility which has been thoroughly discussed
working within simplified toy models.

Systems subject to long range couplings also display unexpected dynamical
features. Starting from out-of-equilibrium initial conditions they are
occasionally trapped in long lasting regimes, termed Quasi Stationary States
(QSS), whose lifetime diverges with the number of elements, $N$, belonging to the
system under scrutiny \cite{Yamaguchi}.

The QSSs have been shown to relate to the stable steady states of the Vlasov
equation, which governs the dynamical evolution of the single particle
distribution function in the continuum limit $N \rightarrow \infty$
\cite{reviewRuffo,Yamaguchi,AntoniazziPRL1,fel}. Working within this setting, one can implement an analytical
procedure, fully justified from first principles, to clarify some aspects of QSS
emergence. The idea, inspired to the seminal work of Lynden-Bell
\cite{LyndenBell68}, is based on the definition of a locally-averaged
(``coarse-grained") distribution function, which translates into an entropy
functional, as follows from standard statistical mechanics prescriptions. By
maximizing such an entropy, while imposing the constraints of the dynamics,
returns a closed analytical expression for the single particle distribution
function of the system in its QSS regime. The predictive adequacy of the
technique was tested versus numerical simulations for specific applications
relevant in e.g. plasma physics, and for the Hamiltonian Mean Field
(HMF) model \cite{antoni-95}, to which we will make extensive reference in the
following. Furthermore, the Lynden-Bell approach allows one to successfully
identify out-of-equilibrium phase transitions separating homogeneous and non
homogeneous steady states \cite{Chavanis_phasetransition,AntoniazziPRL2}. More recently, 
the Lynden-Bell procedure was applied to
an open HMF system, modified by the inclusion of an externally imposed field,
prognosticating the existence of regions of negative susceptibility which were
then observed in direct simulations of the discrete Hamiltonian
\cite{DeninnoFanelliEPL}.

In this paper we revisit the Lynden-Bell analysis for the HMF model. The
theoretical scenario is tested versus Vlasov based simulations, returning an
overall good agreement. We discuss also the impact of the
choice of a monitored quantity on the characterization of the order of the
phase transition in absence of the external field.
The role of the externally imposed field is
assessed, with emphasis on the modification of the Lynden-Bell transition. 
The response of the system to the external forcing results in a smoothing
of the transition that separates homogeneous and non homogeneous regimes, an
observation which a posteriori supports the identification of such phenomenon
with a genuine phase transition. We here anticipate that regions with negative
susceptibility will be also identified in agreement with the Lynden-Bell
scenario depicted in \cite{DeninnoFanelliEPL}. 

The paper is organized as follows. The next section is devoted to introducing
the HMF model and to discussing its continuous analogue. In Section \ref{sec:maxentropy} we
present the Lynden-Bell calculation, revisiting the results with reference to
the unforced system. In Section \ref{sec:magnetization} we present the results
of the numerical simulations, based on a Vlasov code, aimed at verifying
Lynden-Bell's prediction of the presence of an out-of-equilibrium phase transition
in the HMF model. The effect of applying an external magnetic field $h$ is discussed
in Section \ref{sec:field}. Finally, in Section \ref{sec:conclusions}  we sum up and conclude.

\section{The Hamiltonian Mean Field model}
\label{sec:HMF}

The Hamiltonian Mean-Field (HMF) model \cite{antoni-95} describes the
motion of $N$ classical rotors coupled through a mean-field interaction. The system, in 
its standard formulation, can be straightforwardly
modified to include an external perturbation that acts on the particles as a
magnetic-like field \cite{Chavanis_h}. The model is mathematically defined by the following
Hamiltonian:
\begin{equation}
\label{eq:ham}
H = \frac{1}{2} \sum_{j=1}^N p_j^2 + \frac{1}{2 N} \sum_{i,j=1}^N
\left[1 -  \cos(\theta_j-\theta_i) \right] - h\sum_{j=1}^N\cos(\theta_j), 
\end{equation}
where $\theta_j$ represents the orientation of the $j$-th rotor and $p_j$ is its
angular momentum. The scalar parameter $h$ measures the strength of the
magnetic field. Hamiltonian (\ref{eq:ham}) with $h=0$ has been widely studied in
the past as a prototype model of long-range interacting systems. To monitor the
dynamics of the systems, one often refers to the magnetization, a collective
variable defined as 
\begin{equation}
{\mathbf m}=\frac{1}{N} \sum_{i=1}^N (\cos \theta_i, \sin \theta_i)=(m_x,m_y)~.
\end{equation}
The modulus of ${\mathbf m}$, $m=\sqrt{m_x^2+m_y^2}$, measures the degree of bunching of the rotors 
along a given direction. The model can be also interpreted as describing $N$ particles
moving on a circle. Within this interpretation, magnetized regimes signal the presence 
of a localized agglomeration of particles on the circle.

As previously reported in the literature \cite{Yamaguchi}, starting from an
out-of-equilibrium initial condition, the system gets trapped in long lasting
QSSs, whose macroscopic characteristics differ significantly from those
associated to the corresponding equilibrium configurations. QSSs develop for
both $h=0$ and $h \ne 0$ settings.
In the $N \to \infty$ limit, the system is indefinitely stuck in the QSS
phase. 

On the other hand, when performing the limit for infinite system size,
the discrete model Hamiltonian (\ref{eq:ham}) admits a rigorous continuous
analogue. This is the Vlasov equation which governs the evolution of the single
particle distribution function $f(\theta, p , t)$: 
\begin{eqnarray}
  \label{eq:vlasov}
  \frac{\partial f}{\partial t} &+& p\, \frac{\partial f}{\partial \theta} - \frac{dV[f]}{d\theta} 
  \frac{\partial f}{\partial p} = 0\; , \cr
    & & \cr
  V[f](\theta) &=& 1 - \left(m_x[f]+h\right) \cos\theta - m_y[f] \sin\theta\; , \cr
  m_x[f] &=& \int d\theta\, dp\, f \cos\theta \; ,\cr
  m_y[f] &=& \int d\theta\, dp\, f \sin\theta\; ,
\end{eqnarray}
where $V$ is the interaction potential that depends self-consistently on $f(\theta,p,t)$.
According to this kinetic
picture the free streaming of the particles is opposed by a potential term $V[f]$,
reminiscent of the discrete formulation, expressed as a self-consistent function
of the dynamically varying distribution $f(\theta, p , t)$.

In light of the above, the QSSs have been interpreted as stable steady states
of the underlying Vlasov equation. Working within this setting, and invoking the
aforementioned Lynden--Bell violent relaxation theory \cite{LyndenBell68}, one
can progress analytically at least for a simplified choice of the initial
condition.  A short account of the technicalities is provided in the following
section.

\section{The maximum entropy solution}
\label{sec:maxentropy}

Assume the particles to be confined within a bounded domain of phase space,
therein displaying a uniform probability distribution. Label $f_0$ the constant
value of $f(\theta,p,t)$ within the selected domain, as imposed by the
normalization condition. This working ansatz corresponds to dealing with the
``waterbag'' distribution:
\begin{equation}\label{f_0-HMF}
f(\theta,p,0)= \left\{ \begin{array}{ll}
f_0=\frac{1}{4 \Delta \theta \Delta p} & \textrm{if } -\Delta p < p < \Delta p \\
  &\textrm{and } - \Delta \theta < \theta < \Delta\theta \\
0 & \textrm{otherwise,}\\
\end{array} \right.
\end{equation}
that is even in both $\theta$ and $p$: $f(-\theta,-p,0)=f(\theta,p,0)$. For distributions 
endowed with this symmetry, it can be straightforwardly proven that, being $m_y=0$ initially, its 
value remains zero during time evolution. This in turn
implies that also the total momentum $P=\int p f(\theta,p,t) d \theta d p$, which is zero
initially, remains zero during the whole time evolution, i.e. there is
no global rotation of the particles on the circle.
With this choice, one parametrizes the initial condition in terms of the energy
density $u=H/N$ and the initial magnetization ${\mathbf m}=(m_0,0)$. Momentum $P$
cannot be considered as a global invariant, because the presence of an external
magnetic field breaks the translation symmetry $\theta \to \theta + \alpha$. However, for the
initial distributions (\ref{f_0-HMF}), momentum is fixed to zero. 

Under the Vlasov evolution, the waterbag gets distorted and filamented at
smaller scales, while preserving its surface in phase space.  The distribution
stays two-level ($0,f_0$) as time progresses. By performing a local average of
$f$ inside a given mesoscopic box, one gets a coarse-grained profile which is
hereafter labelled $\bar{f}$. As opposed to $f$, the locally averaged function
$\bar{f}$ converges to an asymptotic equilibrium solution which can be
explicitly evaluated via a rigorous statistical mechanics procedure, adapted
from the pioneering analysis of Lynden-Bell. An entropy functional $s(\bar{f})$,
can be in fact associated to $\bar{f}$, through a direct combinatorial
calculation \cite{LyndenBell68}. In the two-level waterbag scenario, the mixing
entropy density $s$ takes the form:
\begin{equation}
\label{entropy_shape}
s[\bar{f}]=-\int \!\!{\mathrm d}p{\mathrm d}\theta \,
\left[\frac{\bar{f}}{f_0} \ln \frac{\bar{f}}{f_0}
+\left(1-\frac{\bar{f}}{f_0}\right)\ln
\left(1-\frac{\bar{f}}{f_0}\right)\right].
\end{equation}
The energy density 
\begin{equation}
u[\bar{f}] = \iint {\mathrm d}\theta {\mathrm d}p \frac{p^2}{2} f(\theta,p,t)+\frac{1-m_x^2-m_y^2}{2}- h m_x
\end{equation}
is conserved. In addition, the normalization of the distribution $\bar{f}$ has to be 
imposed, which physically corresponds to assume constant mass. 
Requiring the entropy to be stationary, while imposing the conservation of energy and mass, 
result in a variational problem that admits the following solution:
\begin{multline}
  \label{eq:barf}
  \bar{f}_{\text{QSS}}(\theta,p)= \\
  \frac{f_0}{ 1+e^{\displaystyle\beta f_0 (p^2/2 -{m}_x \cos \theta -
  {m}_y \sin \theta -h\cos\theta)+\alpha}},
\end{multline} 
where $\alpha$ and $\beta$ are Lagrange multipliers associated, respectively, to mass and energy conservation and $m_x$ and $m_y$ depend on $\bar{f}_{\text{QSS}}$.
The self-consistent nature of Eq.~(\ref{eq:barf}) is evident: $m_x[\bar{f}_{\text{QSS}}]$ and $m_y[\bar{f}_{\text{QSS}}]$ 
are functionals of $\bar{f}_{\text{QSS}}$ and both enter in the determination of  $\bar{f}_{\text{QSS}}$ itself.
We also emphasize that the label $\text{QSS}$ is introduced to recall that the stationary solution of the Vlasov 
equation are indeed associated to QSSs of the discrete $N$-body dynamics.

Back to solution (\ref{eq:barf}), one can determine the predicted values of $m_x$, $m_y$, $\alpha$ and $\beta$  
once the energy $e$, the field $h$ and the waterbag height $f_0$ are being assigned. This step is performed numerically, at 
sought accuracy, via a Newton-Raphson method. 

\begin{figure}[htbp]
  \centering
  \includegraphics[width=\linewidth]{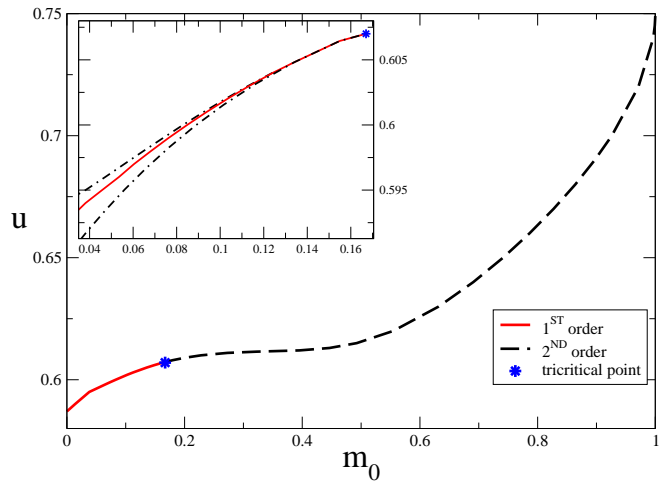}
  \caption{Phase diagram in the $(m_{0},u)$ plane at $h=0$. The full line refers to the the first order
  transition, while the dashed line stands for the second order one. The symbol traces the exact 
  location of the tricritical point. Inset: lateral edges of the coexistence regions in the first order 
  region, as predicted by the theory.
} 
  \label{fig:phase-diagram}
\end{figure}

Consider first the limiting case $h=0$. Depending on the value of the predicted magnetization $m$, one can ideally 
identify two different regimes: the homogeneous case corresponds to  $m=0$ (non-magnetized), while the non-homogeneous 
setting is found for (magnetized) $m \ne 0$ solutions. A phase transition \cite{Chavanis_phasetransition,AntoniazziPRL1}  
materializes in the parameters plane $(m_0,u)$ and the resulting scenario is depicted in Fig.~\ref{fig:phase-diagram}.
When fixing the initial magnetization and decreasing the energy density, the system  passes from homogeneous to 
non-homogeneous QSS. The parameters plane can be then formally partitioned into two zones    
respectively associated to an ordered non-homogeneous phase, $m \ne 0$, 
(lower part of Fig. \ref{fig:phase-diagram}), and a disordered homogeneous state, $m=0$ (upper part).
These regions are delimited by a transition line,  collection of all the critical points ($m_0^c,u^c$), 
which can be in turn segmented into two distinct parts.

The full line stands for a first order phase transition:
the magnetization experiences a finite jump  when crossing the critical value ($m_0^c,u^c$).
Conversely, the dashed line refers to a second order phase transition: 
the magnetization is continuously modulated, from zero to positive values,
when passing the curve from top to bottom. First and second lines merge in a tricritical point.
 
The case with $h \ne 0$ has been recently addressed in \cite{DeninnoFanelliEPL} for what
concerns the Lynden-Bell theory and working at constant $f_0$, while the equilibrium properties have been thoroughly
studied in \cite{Chavanis_h}. 
Again, the Lynden Bell approach proves accurate in predicting the macroscopic behavior as seen in the $N$-body simulations. 
Interestingly,  below a threshold in energy the system shows negative susceptibility $\chi = \partial M / \partial h$, 
the magnetization decreasing when the strength of $h$ is enhanced. Conversely, above the critical energy value, the 
magnetization amount grows with $h$, which corresponds to dealing with positive susceptibility. Besides providing 
an a posteriori evidence on the adequacy of the Lynden-Bell technique,
the presence of a region with negative susceptibility, was interpreted in \cite{DeninnoFanelliEPL} as the signature 
of an out-of-equilibrium ensemble inequivalence. Furthermore, the presence of the field $h$ removes the phase
transition and the magnetization continuously decreases from unity, at zero temperature, to zero, at infinite
temperature. Therefore, a modest, though non negligible spatial polarization of the rotors is present also 
in the parameters region that was destined to homogeneous phases in the limiting case $h=0$.

Starting from this setting, we have decided to perform a campaign of Vlasov based simulations to challenge the rich 
scenario predicted within the realm of the Lynden-Bell violent relaxation theory. By numerically solving the Vlasov 
equation, we avoid dealing with finite size effects, as stemming in direct $N$-body schemes, and so provide a 
more reliable assessment of the overall correctness of the theory. The results of the investigations are reported 
in the forthcoming sections.

\section{Magnetization and its fluctuations}
\label{sec:magnetization}
The Vlasov equation (\ref{eq:vlasov}) can be resolved numerically. To this end, we use the
semi-Lagrangian method with cubic spline interpolation, as implemented in the
{\tt vmf90} program that has been used already in Ref.~\cite{de_buyl_cnsns_2010}
with the HMF model.

In order to study the properties of the QSS regime, we adopt the following
procedure:
\begin{enumerate}
\item The system is started with a waterbag initial condition (\ref{f_0-HMF}).
\item It is run without collecting data between times $t_0=0$ and $t_1=100$.
\item Time averages of the magnetization $m$, of $m_x$ and of the variance of the
magnetization $m$ are performed in the time range between $t_1=100$ and $t_2=200$,
defining
\begin{equation}
\overline{m}=\frac{1}{t_2-t_1} \int_{t_1}^{t_2} m(t) dt~,
\end{equation}
\begin{equation}
\overline{m}_x=\frac{1}{t_2-t_1} \int_{t_1}^{t_2} m_x(t) dt~,
\end{equation}
\begin{equation}
\sigma_m^2=\frac{1}{t_2-t_1} \int_{t_1}^{t_2} (m(t)-\overline{m})^2 dt~.
\end{equation}
\end{enumerate}
We thus skip the strong oscillations of the initial violent relaxation and focus
on the subsequent dynamical regime, where, anyway, some oscillations are
still present, which we quantify by the standard deviation $\sigma_m$.
We repeat the above procedure on a grid of 39 by 39 points in the $(m_0,u)$ plane, 
each one corresponding to a simulation.
Performing such a study allows us to assess in a systematic manner the behaviour of the average
magnetization in the QSS regime and to compare numerical results with
Lynden-Bell's theory. Whether the theory is flexible enough to accommodate
all the general features of the resulting diagram depends on it taking into
account in a comprehensive manner the behaviour of the model.

\begin{figure}[h]
  \centering
  \includegraphics[width=\linewidth]{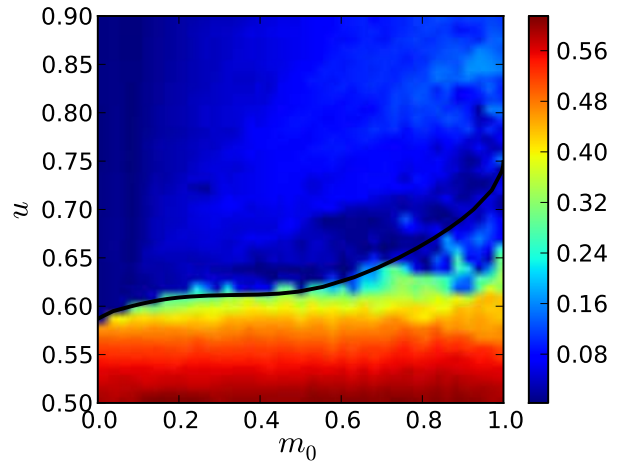}
  \caption{Average magnetization $\bar m$ for the HMF model with no external
    field. The transition predicted by Lynden-Bell's theory is indicated by the
    full black line.}
  \label{fig:nofield}
\end{figure}

\begin{figure}[h]
  \centering
  \includegraphics[width=\linewidth]{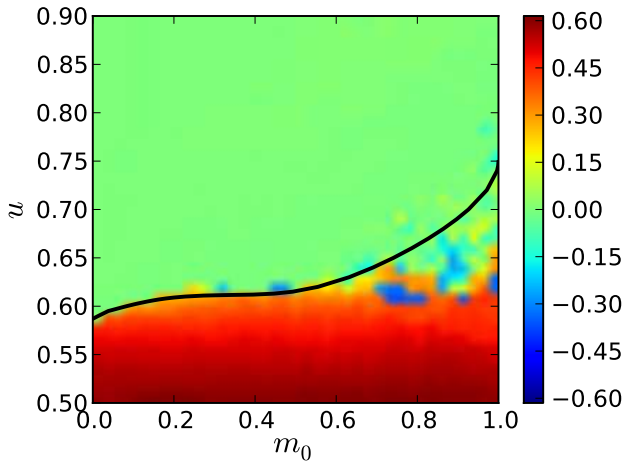}
  \caption{Average of the $x$-component of the magnetization $\overline{m}_x$ for the HMF 
  model with no external field. The transition predicted by Lynden-Bell's theory 
  is indicated by the full black line.}
  \label{fig:nofield_mx}
\end{figure}

\begin{figure}[h]
  \centering
  \includegraphics[width=\linewidth]{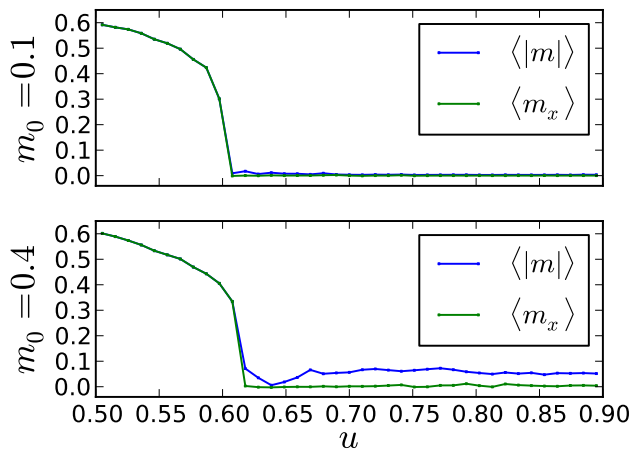}
  \caption{Comparison of $\overline{m}$ and $\overline{m}_x$ as
  a function of $u$ for $m_0=0.1$ and $m_0=0.4$.}
\label{fig:MxM}
\end{figure}

The average value of the magnetization taken from Vlasov simulations is
displayed in Fig.~\ref{fig:nofield}. The line of transition provided by Lynden-Bell's
theory is displayed on top and we observe that it separates satisfactorily the
region $m>0$ from the region $m \approx 0$ for $m_0 \lesssim 0.6$. The transition is sharp for low
values of $m_0$, corresponding to the prediction of Lynden-Bell's theory that the
transition is of first order. The transition is smoother for larger values 
of $m_0$, corresponding to the Lynden-Bell's prediction of a second order transition.
We display in Fig.~\ref{fig:nofield_mx} the same diagram as the one of
Fig.~\ref{fig:nofield}, but using $\overline{m}_x$ instead of $\overline{m}$. 
The general aspect of the diagram is similar, although, as remarked
in \cite{pakter_levin_prl_2011}, the transition looks overall sharper when 
using $\overline{m}_x$ as an order parameter. We note that Lynden-Bell's transition line 
separates very well the non-homogeneous from the homogeneous phase for
$m_0 \lesssim 0.4$. For higher values of $m_0$ there are simulations for which
$\overline{m}_x < 0$ (blue spots below the Lynden-Bell's transition line
in Fig.~\ref{fig:nofield_mx}). This occurs when the phase of ${\mathbf m}$ is
$\pi$ (instead of zero). The phase could in principle take any value in
$[-\pi,\pi]$, but since $m_y=0$ the only dynamically accessible values 
are $0$ and $\pi$. The fact that $m_x$ can flip from positive to negative
values is also an indication of the presence of a second-order phase transition. 
Indeed, these flips are not present in the first order phase transition region,
where an entropic barrier at the phase transition separates positive and
negative values of $m_x$.
The difference in the values of $\overline{m}$ and $\overline{m}_x$ can only
arise from time fluctuations of $m_x(t)$. Indeed
\begin{equation}
\overline{m^2}=\overline{m_x^2}=\overline{m_x}^2+\sigma_m^2
\end{equation}
To illustrate the difference between these two quantities, 
we compare them in Fig.~\ref{fig:MxM} for $m_0=0.1$ and $0.4$.
In the low energy phase the two quantities are indistinguishable,
prooving that the fluctuations are small. It is confirmed that
$\overline{m}_x$ goes sharply to zero at the transitions energy and
remains zero in the whole high energy phase, as found in \cite{pakter_levin_prl_2011}. On the contrary,
$\overline{m}$, the quantity measured in \cite{AntoniazziPRL2}, has a tail of positive values at high energy, especially
visible for $m_0=0.4$, proving that fluctuations are here larger.

\begin{figure}[h]
  \centering
  \includegraphics[width=\linewidth]{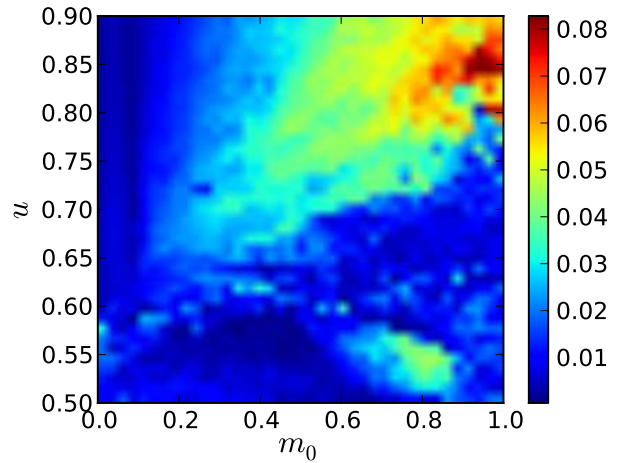}
  \caption{Amplitude of oscillations measured by $\sigma_m$ for the HMF
    model with no external field.}
  \label{fig:nofield_stddev}
\end{figure}

The variance of the magnetization, $\sigma_m$, is displayed in Fig.~\ref{fig:nofield_stddev}. 
It is confirmed that, below the transition line predicted by Lynden-Bell's theory, fluctuations
of $m$ are small. They are instead large in the high energy region above the Lynden-Bell's
transition line for $m_0 > 0.4$.

By pointing out the different results that arise from the choice of
different order parameters, $\overline{m}$ or $\overline{m}_x$, we hope that 
in future studies the problem of the out-of-equilibrium phase transition in
the HMF model will be analyzed more carefully.

\section{Response to the application of a small magnetic field}
\label{sec:field}

\begin{figure}[h!]
  \centering
  \includegraphics[width=\linewidth]{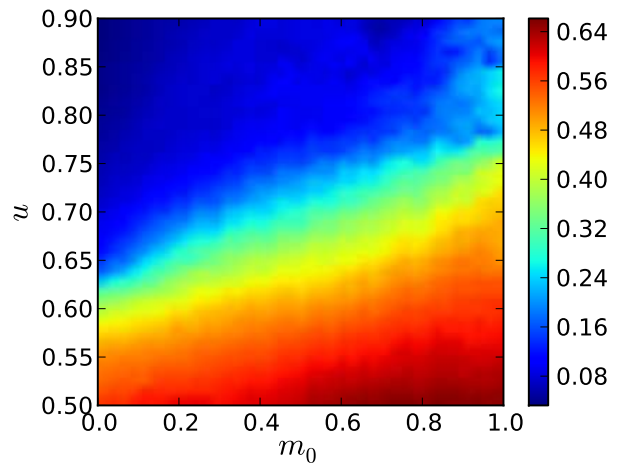}
  \caption{Average magnetization $\overline{m}$ for the HMF model with an external
    field $h=0.1$.}
  \label{fig:field}
\end{figure}

In this Section, we present the results of simulations for the HMF model with a
small external magnetic field, we choose $h=0.1$. The average value of the magnetization 
$\overline{m}$ obtained from Vlasov simulations is displayed in Fig.~\ref{fig:field}.
The phase transition is removed by the application of the field, as it happens for
equilibrium phase transitions. The magnetization, for all values of $m_0$, decreases
smoothly to zero as the energy is increased.

\begin{figure}[h!]
  \centering
  \includegraphics[width=\linewidth]{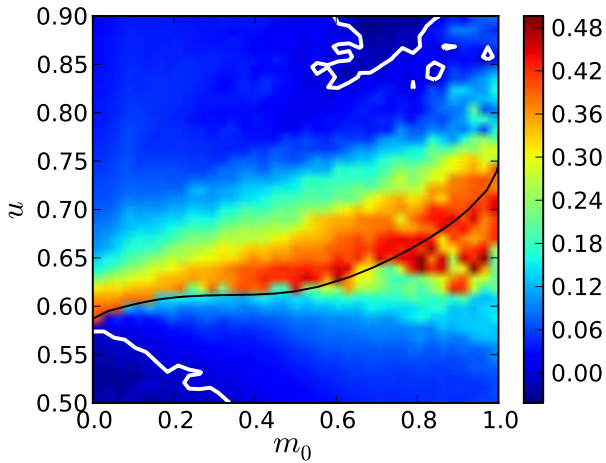}
  \caption{Difference of average magnetization between simulations with $h=0.1$
    and simulations with $h=0$. The white lines indicate the zero level, so that
    the darker region close to ($m_0=0$,$u=1/2$) is a region of negative magnetic
    susceptibility.}
  \label{fig:diff}
\end{figure}

Magnetic susceptibility $\chi = \partial m / \partial h$
characterizes the response of the system to the application of an external field. It has been
shown in Ref.~\cite{DeninnoFanelliEPL} that certain parameter regions display a
negative magnetic susceptibility. This is a signature of ensemble inequivalence,
shown here in a {\em out-of-equilibrium} setting, as the system is trapped in
the QSS regime and does not reach equilibrium.
In this Section, we provide a similar measure by taking the difference of
the average magnetization between simulations with $h=0.1$ and simulations with
$h=0$. The result is displayed in Fig.~\ref{fig:diff}.
While our computations provide a discrete difference instead of a derivative,
obtaining a lower value of the average magnetization for $h=0.1$ than for $h=0$
is the sign of a negative susceptibility nonetheless. As expected from
Ref.~\cite{DeninnoFanelliEPL}, a region of Fig.~\ref{fig:diff} displays $\chi <0$ for low values 
of $m_0$, in the vicinity of the first-order transition found
in the theory. We are thus able to confirm the theoretical prediction on the
basis of Vlasov simulations.
Figure~\ref{fig:diff} also displays a large value of $\chi$ around the
transition line predicted by Lynden-Bell's theory.

\section{Conclusions}
\label{sec:conclusions}

We have performed a study on the adequateness of Lynden-Bell's theory
as compared to numerical simulations of the Vlasov equation for the Hamiltonian
Mean-Field model. Our results confirm previous studies based on $N$-body
simulations on the general quality of the phase diagram. We extended the
knowledge of the phase diagram by several additional measurements: the amplitude
of oscillations, $\sigma_m$, the magnetic susceptibility. By doing so, we point out that in regions where
non negligible fluctuations of the magnetization $m$ occur, the theory is not expected to work, whereas the
agreement is quite good between numerical simulations and theory in regions
where fluctuations are small. We also confirmed that there are regions of negative 
susceptibility, as predicted in Ref.~\cite{DeninnoFanelliEPL}.

Finally, we also discuss the more fundamental, but related issue, of the
appropriate thermodynamical quantity to follow in the simulations. This latter
issue is not touched upon in the literature but reveals qualitatively different
results for the transition from magnetized to non-magnetized regimes. As of now,
Lynden-Bell's theory provides a clear determination of the order of the
transition and a tricritical point is found in the phase diagram. However, the
dynamical aspects of the transitions are not yet elucidated, as is known from
numerical simulations, even if steps are taken in that
direction~\cite{pakter_levin_prl_2011,de_buyl_et_al_rsta_2011,de_buyl_et_al_in_prep}.

\section{Acknowledgements}
D.F. thanks Giovanni De Ninno for discussions. S. R. acknowledges support of the contract LORIS
(ANR-10-CEXC-010-01).


\begin{thebibliography}{99}

\bibitem{reviewRuffo} A. Campa, T. Dauxois and S. Ruffo, Phys. Rep. {\bf 480}, 57 (2009).
\bibitem{Leshouches} T. Dauxois, S. Ruffo and L. Cugliandolo (Eds.), 
{\it Long-Range Interacting Systems}, {\it Lecture Notes of the Les Houches Summer School: 
Volume 90, August 2008}, Oxford University Press (2009).
\bibitem{Yamaguchi} Y.Y. Yamaguchi, J. Barr{\'e}, F. Bouchet, T. Dauxois and S. Ruffo, Physica A {\bf 337}, 
36 (2004).
\bibitem{AntoniazziPRL1}   A. Antoniazzi, F. Califano, D. Fanelli and S. Ruffo,
Phys. Rev. Lett. {\bf 98}, 150602 (2007).
\bibitem{fel}  J. Barr{\'e}, T. Dauxois, G. De Ninno, D. Fanelli and S. Ruffo,
Phys. Rev. E {\bf 69}, 045501(R) (2004).
\bibitem{LyndenBell68} D. Lynden-Bell and R. Wood, Mon. Not. R. Astron. Soc.
{\bf 138}, 495 (1968).
\bibitem{antoni-95} M. Antoni and S. Ruffo, Phys. Rev. E {\bf 52}, 2361-2374 (1995).
\bibitem{Chavanis_phasetransition} P. H. Chavanis, Eur. Phys. J. B {\bf 53}, 487 (2006).
\bibitem{AntoniazziPRL2} A. Antoniazzi, D. Fanelli, S. Ruffo, Y.Y. Yamaguchi, Phys. Rev. Lett.
{\bf 99}, 040601 (2007).
\bibitem{DeninnoFanelliEPL} G. De Ninno and D. Fanelli, Europhys. Lett. in press arXiv:1011.2981 (2011). 
\bibitem{Chavanis_h} P. H. Chavanis, Eur. Phys. J. B {\bf 80}, 275 (2011). 
\bibitem{de_buyl_cnsns_2010} P. de Buyl,Commun. Nonlinear
  Sci. Numer. Simulat. {\bf 15}, 2133-2139 (2010).
\bibitem{pakter_levin_prl_2011} R. Pakter and Y. Levin, Phys. Rev. Lett. {\bf 106}, 200603 (2011).
\bibitem{de_buyl_et_al_rsta_2011}
P. de~Buyl, D. Mukamel and S. Ruffo, Phil. Trans. R. Soc. A, {\bf 369}, 439 (2011).
\bibitem{de_buyl_et_al_in_prep}
P. de Buyl, D. Mukamel and S. Ruffo, arXiv:1012.2594 (2010).

\end {thebibliography}
\end{document}